# The TOF method for the LSDS-100 spectrometer

A.A. Alekseev[1], Yu.V. Grigoriev[1,2], V.A. Dulin[1,3], O.N. Libanova[1], V.L. Matushko[1], Zh.V. Mezentseva[2], A.V. Novikov-Borodin[1], Yu.V. Ryabov[1]

[1]*Institute for Nuclear Research RAS, Moscow, Russia*
[2]*Joint Institute for Nuclear Research, Dubna, Russia*
[3]*Institute of Physics and Power Engineering, Obninsk, Russia*

The first lead neutron slowing-down spectrometer (LSDS) from a pulsed source in elastic scattering on lead nuclei has been constructed in the laboratory of the atomic nucleus of the Physics Institute of the Russian Academy of Sciences LPI [1,2,3]. Currently, there are several operating lead neutron slowing-down spectrometers: LSDS-100 in Russia [4], RINS in United States [5], KULS in Japan [6] and others. A relation between an energy of lead slowing-down moderated neutrons $E$ (eV) and the time delay of $t$ (μs) is described by the expression: $E(t) = K / (t + t_0)^2$, where values of $K=170.5$ (keV×μs$^2$), $t_0 = 0.3$ (μs) in the case of the LSDS-100 have been measured in the previous experiment [4]. The energy resolution of LSDS is low (30-45%) and is being determined experimentally, but the aperture ratio luminosity of the neutron flux detector is $10^2$-$10^4$ times greater than in the case of a time of flight method technique. It allows to make experiments with small amount of substance and small cross-sections of its interaction with the slowed moderated neutrons.

To expand the experimental possibilities of LSDS-100 with total weight of 100 tons, which is mounted in the form of a parallelepiped of 3348×1620×1728 mm$^3$ dimensions from lead prisms of 1 and 0.25 ton weights and of 99,996 % natural lead purity, it is proposed to install the polyethylene moderator in the form of a plate of 100×100×30 mm$^3$ dimensions or of a disk with 100 mm diameter and a thickness of 30 mm in one of available vertical channel-wells over a lead target. Such moderator forms a wide spectrum of neutrons from thermal to fast 200 MeV neutrons. If the moderator is cooled up to the temperature of liquid nitrogen, then the thermal part of neutrons will be shifted to the range of cold neutrons. To improve background conditions and to provide the neutron beam forming, it is necessary to mount inside the vertical channel-well over the moderator the neutron guide, which is the vacuum pipe of 90 mm diameter and of 1-10 m length with the collimators nearby the moderator and at the end of the neutron guide in front of the detector. It is supposed to use as detectors the fast counters of neutrons, γ-rays and the fission chambers. Certainly, due to the moderator the mounting of the proposed equipment of the TOF-spectrometer in the lead body of the LSDS-100 will slightly worsen its energy resolution, but in fact the spectrometer geometry is not changed. The parameters of the LSDS-100 combined with the TOF-spectrometer will be defined experimentally. If the mutual influence of spectrometers is negligible, it is possible to make measurements simultaneously on both facilities. Otherwise the moderator should be removed from the body of the LSDS-100. The TOF-spectrometer may be used also when the LSDS-100 is running. In that case it is necessary to install the good detector shielding from the background neutrons leaking from a body of a lead cube.

With using the TOF-technique, the energy $E(t)$ of neutrons, registered by the detector, is defined by the equation: $E(t) = mv^2/2 = (72.3·L/t)^2$, where $t$(μs) is a time of flight of a neutron of distance $L$(m) from the moderator to the detector. The energy resolution is defined from the following ratio: $\Delta E/E=2v/L=0.0276\sqrt{E}\Delta t/L$, where $v$ is a speed of a neutron with an energy $E$, $\Delta E$ is an uncertainty of energy, $\Delta t$ is an uncertainty of time.

The values of the energy resolution of the time-of-flight spectrometer at various energies and flying bases at $\Delta t=2$ μs are presented in Table 1.

**Table 1.** The resolution $\Delta E/E$ (%) versus the energy of neutrons and the length of the flying base $L$.

| $E$ (keV) / $L$ (m) | 0.9 | 1.6 | 2.5 | 49 |
|---|---|---|---|---|
| 1 | 1.67 | 2.01 | 2.74 | 3.66 |
| 2 | 0.83 | 1.01 | 1.37 | 1.66 |
| 4 | 0.41 | 0.50 | 0.69 | 0.96 |
| 8 | 0.21 | 0.26 | 0.34 | 0.48 |
| 50 | 0.033 | 0.040 | 0.054 | 0.060 |

It is seen from the Table 1 that the energy resolution of the TOF- spectrometer even on the shortest base of 1 m at the energies low than 50 keV is better than of the LSDS-100. As soon as the integral output of the fast neutrons in the lead target of LSDS-100 is approximately $10^{14}$ n/(cm$^2\times$s) at the average proton current of 5 μA, the integral neutron flow on the neutron absorbing detector surface with square of 100 cm$^2$ at the distance of 1 m from the pulsed neutron source will be approximately $10^{10}$ n/(cm$^2\times$s) and at the distance of 10 m will be $10^8$ n/(cm$^2\times$s), which is 100 times higher the flow density at the middle of the each of five experimental channels of LSDS-100 [4]. It is worth noting that besides traditional researches on the fundamental and applied nuclear physics the proposed combined spectrometer based on LSDS-100 and TOF- spectrometer may also be used for modeling the different types of targets and to investigate the isomeric states excited by protons in wide range of energies. As soon as there are five experimental cylindrical channels with 65-80 mm diameter, two wells with 100×100 mm$^2$ cross section square and the thermal graphite prism in the body of LSDS-100, it allows to make various experiments on development of techniques of a transmutation of a long-living waste of atomic power engineering.

In order to study the neutron fission cross-sections of minor actinides it is supposed to use the fast chambers of IPPE (Institute of Physics and Power Engineering) with the parameters presented in Appendix.

In conclusion the authors would like to thank the management of experimental physics and accelerator departments and of INR RAS for support of this work.

# Appendix

The parameters of new ionized fission chambers with thin layers of minor actinides, which are supposed to use for measurements at the LSDS-100 by the transmutation program.

### The detection of velocities of the fission reactions

There are few known methods of determination of the ratio of velocities of isotopes fission reactions, which are interesting for the experimental physics of reactors. The mostly widespread one of them is the method of calibration in thermal column, which is easily realized on test stands with such column and the cross-sections of isotopes fission in the thermal column are well-known.

Another method is the technique of determination of the ratio of the average fission cross-sections with help of the absolute fission chambers with known number of nuclei in the layer and known efficiency of the registration of the fission fragments. It is suitable for determination of the velocities of the fission reactions of the most of trans-actinides, having threshold character, and also for determination of absolute fission velocities of $^{239}$Pu and $^{252}$Cf.

## The absolute plane fission chambers

To calculate the ratio of the average isotope fission cross-section for the technique, where the absolute plane fission chambers of the test stand (see Fig.1) are used during measurements, there was used the expression (1):

$$\frac{\sigma_f^i}{\sigma_f^j} = \sigma_f^{i/j} = \frac{N^i \varepsilon^j m^j}{N^j \varepsilon^i m^i} - \sum_k^n \left( a^{k/j} \cdot \sigma_f^{k/j} \right), \quad (1)$$

where

$N^i$, $N^j$ are the count velocity of the fission chamber with the layer of the $i$-th nuclide and the count velocity of the fission chamber with the layer of the $j$-th nuclide in the active zone;

$\varepsilon^i$, $\varepsilon^j$ are the efficiencies of the fission fragments registration by the fission chamber with the layer of $i$-th nuclide and by the fission chamber with the layer of $j$-th nuclide correspondingly;

$m^i$, $m^j$ are the number of nuclei in the layer of $i$-th nuclide and $j$-th nuclide correspondingly;

$i, j$ are the atomic number of being measured and reference nuclide correspondingly;

$k$ are the atomic numbers of doped nuclides, containing in the active layer of the measured ($i$-th) nuclide;

$a^{k/j}$ is the ratio of the number of nuclei of doped nuclides to the number of nuclei of the basic nuclide (being measured), $\sigma_f^{k/j}$ is the ratio of the average fission cross-sections of doped nuclides to the number of nuclei of basic nuclide. It is supposed here for simplicity that the contaminants in the layer of the basic nuclide (it is usually $^{239}$Pu, see the Table 2) may be neglected.

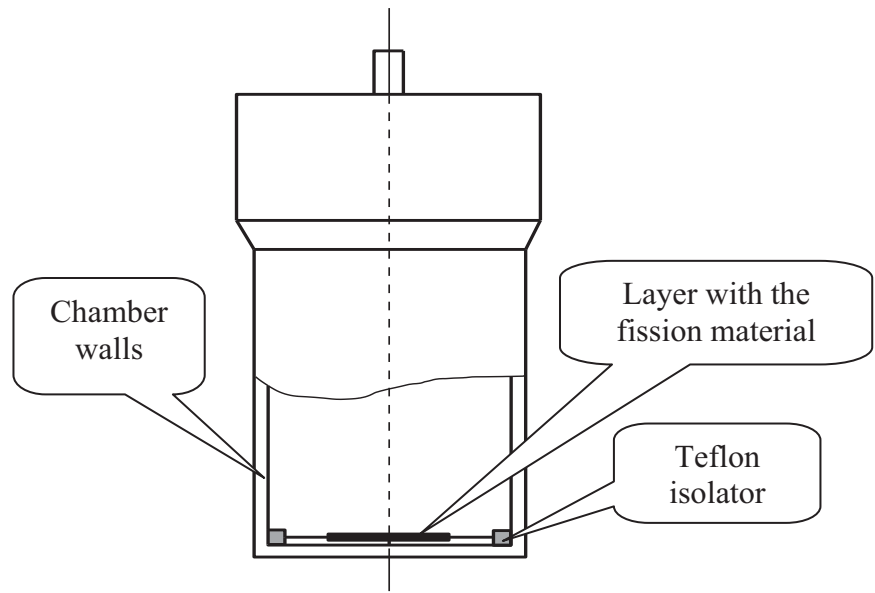

Fig.1 The scheme of the absolute fission chamber

**Table 2**. The rated isotope composition ionization chamber of Institute for Physics and Power Engineering (IPPE).

| Basic element of the layer | Date of test of the isotope composition | Isotopes in the layer | Half-decay period (years) | Isotope composition (mole) | Number of spontaneous fissions (number/sec) |
|---|---|---|---|---|---|
| Pu-238(2) | 6.03.1992 | Pu-238<br>Pu-239<br>Pu-240 | 87.74<br>$2.411\times10^4$<br>$6.537\times10^3$ | $1.8280\times10^{-8}$<br>$0.1894\times10^{-8}$<br>$0.0586\times10^{-8}$ | |
| Cm-244(2) | 20.01.94 | Cm-244<br>Cm-245<br>Cm-246<br>Cm-247<br>Cm-248<br>Pu-240 | 18.11<br>8500<br>$4.73\times10^3$<br><br><br>$6.537\times10^3$ | $1.4450\times10^{-8}$<br>$0.0504\times10^{-8}$<br>$0.1657\times10^{-8}$<br>$0.0052\times10^{-8}$<br>$0.0043\times10^{-8}$<br>$0.0920\times10^{-8}$ | 14.222<br>----------<br>1.221<br>----------<br>0.139<br>---------- |
| Cm-245 | 05.07.12 | Cm-244<br>Cm-245<br>Cm-246 | 18.11<br>8500<br>$4.73\times10^3$ | $0.029\times10^{-8}$<br>$2.87\times10^{-8}$<br>$0.014\times10^{-8}$ | 0.29<br><br>0.1 |
| Am-241 | 25.02.92 | Am-241 | 432.6 | $2.640\times10^{-8}$ | |
| Am-243(2) | 25.02.92 | Am-243<br>Am-241 | $7.38\times10^3$<br>432.6 | $7.160\times10^{-8}$<br>$0.200\times10^{-8}$ | |
| Pu-239-(1) | 20.03.91 | Pu-239<br>Pu-240 | $2.411\times10^4$<br>$6.537\times10^3$ | $19.540\times10^{-8}$<br>$0.030\times10^{-8}$ | |
| Pu-239-1 | 20.03.91 | Pu-239<br>Pu-238<br>Pu-240 | $2.411\times10^4$<br>87.74<br>$6.537\times10^3$ | $41.970\times10^{-8}$<br>$0.026\times10^{-8}$<br>$0.070\times10^{-8}$ | |
| Pu-239-2 | 20.03.91 | Pu-239 | $2.411\times10^4$ | $16.400\times10^{-8}$ | |
| Pu-239-3 | 20.03.91 | Pu-239<br>Pu-240<br>Pu-238 | $2.411\times10^4$<br>$6.537\times10^3$<br>87.74 | $20.750\times10^{-8}$<br>$0.035\times10^{-8}$<br>$0.013\times10^{-8}$ | |
| Pu-240 | 20.03.91 | Pu-240<br>Pu-239<br>Pu-241 | $6.537\times10^3$<br>$2.411\times10^4$<br>14.4 | $15.710\times10^{-8}$<br>$.007\times10^{-8}$<br>$.012\times10^{-8}$ | |
| Np-237 | 20.03.91 | Np-237 | $2.14\times10^6$ | $182.3\times10^{-8}$ | |

## The determination of the amount of the fissionable material

The amount of the nuclei of the fissionable isotope were determined from the measurements of the α-activity of the foil (used as one of the electrode of the chamber) with the deposited layer of the nuclide before the placement it to the chamber. The spectrum of the α-particles has been measured by means of the silicon semiconductor detector. The foil and the detector have been placed into the vacuum with the "good" geometry conditions. Depending on values of α-activity the three solid angles with $1/\Omega = 2273 \pm 12$; $6185 \pm 20$ and $41400 \pm 170$ have been used. Each time the spectrum of the α-particles was measured. The measurements for some foils were made at two solid angles. Since there was the high α-activity from $^{241}$Am and a small infusion of $^{244}$Cm, the spectrum of α-particles for the foil Am-243(2) has been additionally measured by the semiconductor detector with the high resolution.

The producer of the foils (V.G. Khlopin Radium Institute - KRI) has made the measurements with accuracy ± 1 % (1 σ) in 1991 and ± 1.5 % (1 σ) later. The comparisons of results of IPPE and KRI and the ratio IPPE/KRI are presented in Table 3. The discrepancy does not exceed the declared errors. One may consider that the number of nuclei of the basic fissionable isotope in chambers is known not worse than ± 1 % (1 σ).

**Table 3.** The results of the foils' calibration.

| The number of the foil | The layer | The mass of the basic isotope in μg (KRI) | The total α-activity, Bq | | |
|---|---|---|---|---|---|
| | | | KRI | IPPE | IPPE/KRI |
| 3. | Am-243-3 | 1.332 | $3.073\times10^4$ | $3.050\times10^4$ | - 0.8 % |
| 4. | Pu-238 | 3.71 | $2.351\times10^6$ | $2.334\times10^6$ | - 0.8 % |
| 5. | Cm-244 | 3.72 | $1.115\times10^7$ | $1.111\times10^7$ | - 0.4 % |
| 1. | U-233 | 425 | $5.640\times10^5$ | $5.650\times10^5$ | + 0.2 % |
| 2. | Pu-239-1 | 100.3 | $2.309\times10^5$ | $2.286\times10^5$ | - 1.0 % |
| 3. | Pu-239-2 | 39.2 | $1.105\times10^5$ | $1.090\times10^5$ | - 1.5 % |
| 4. | Pu-239-3 | 49.6 | $1.397\times10^5$ | $1.379\times10^5$ | - 1.4 % |
| 5. | Pu-240 | 37.7 | $3.166\times10^5$ | $3.176\times10^5$ | + 0.3 % |
| 6. | Np-237 | 432 | $1.127\times10^4$ | $1.118\times10^4$ | - 0.8 % |
| 7. | Am-241 | 49.1 | $6.221\times10^6$ | $6.220\times10^6$ | 0 % |
| average | | | | | $0.72 \pm 0.16\%$ |
| 1. | Am-243(2) | 17.4 | $1.950\times10^5$ | $1.923\times10^5$ | - 1.5 % |
| 2. | Cm-244(2) | 3.81 | $1.140\times10^7$ | $1.133\times10^7$ | - 0.6 % |
| 3. | Cm-245 | 7.03 | | | $2\pm4\%)*$ |

)* The mass of the basic isotope in μg (IPPE) is equal to $7.15 \pm 0.3$. It is determined by the calibration in the thermal column together with the chamber of Pu-239-3. The fission cross-sections and the Westcott factors in the thermal column have been taken to be equal to 744 barn and 1.062, $2143 \pm 60$ and 0.941.

### The design of the ionization chamber with the layer of Cf

One of the electrodes of the chamber (see Fig.1) is the plate bottom of the cylindrical vessel (glass) (the diameter is 44 mm; the wall thickness is 0.7 mm). At the distance of 1 mm from it there is another electrode in the form of the thin disk. In the center of this disk (from the bottom of the glass) the layer of the spontaneously fissile element $^{252}$Cf is placed as a spot with the diameter of 20 mm. The vessel and the electrode are made from the polished stainless steel. In practice, the bottom of the chamber is convex due to the procedure of the air disposal and argon filling: the chamber is being filled by the argon up to the pressure of 11 atm and the bottom deforms, after that this mixture is released up to 1 atm. The procedure is repeated 10 times, so the part of the rest air is close to $10^{-10}$ atm and then the chamber is filling by argon up to the pressure of 4-5 atm of argon. It has been shown on practice that for such chambers the ratio of amplitudes of α-overlaps and the basic distribution of fragments is minimal.

The last deformation leads to the 'sphericity' of the bottom and exceeding of the gaseous layer may be $(0.2 \div 1)$ mm over the center of the bottom. The operating offset voltage is 400 V. One may suppose that the efficiency of the registration of the fission act (i.e. the registration of the fragment) in the chamber with quite thin layers will be close to unity. However, the necessity to have the amount of fissile elements enough for getting the statistics in the fast neutrons flow ($\sim 10^9$ n/cm$^2$×sec) necessitate to use the more thick layers.

Some part of an energy of fragments releases in the layer (the bigger the deeper the depth and the closer the direction of the fragment outlet to the electrode surface). It leads to losses of some fragments, going out to the gas with energies less the energy from pulses overlaps due to the α-activity of the layers.

Some fragments with trajectory close to the normal to the electrode, may reach the opposite electrode (the bottom), that also leads to decreasing the ionization into the gas and also leads to the decreasing the amplitude of the fragment.

## The determination of the efficiency of the fission act registration

The spectrum of the fragments in the chamber with the distance $d$ between electrodes is resulted as the counting of number and the energy of the fragments, escaping into gas from different depths with different angles to the surface of the layer. The part of the fragments' energy in gas is defined from the geometrical location of their trajectory and the dependence of the fragment energy $E(x)$ on the passing distance $x$:

$$E(x) = E(1 - x/R)^N, \qquad (2)$$

where $R$ is a fragment range, and $N \approx 1.8$. Counting for each such chamber the number and the energy of fragments (i.e. their spectrums) escaping into gas from different depths of the layer and then for different thicknesses of the layers, one can achieve the reasonable correspondence of the calculation and the experiment by the shape for all amplitude range (channels) and, in particular, in the range of the 'visible plateau' between the decrease of α-overlaps and the basic fragments' distribution (50-100 channels). As an example, the measured and calculated spectrums are presented on Fig.2 for the chamber with the layer of $^{252}$Cf.

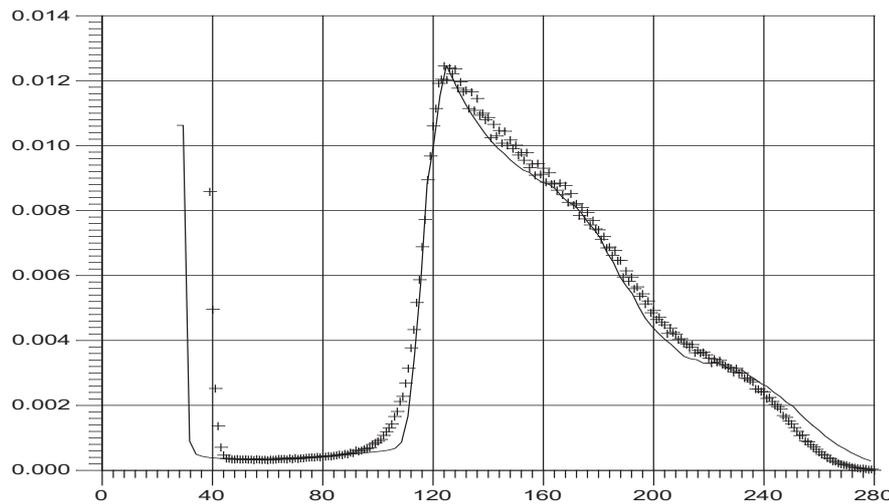

Fig.2 The comparison of the calculated spectrum of fragments (solid line) and the experimental spectrum (+++) for the chamber with the layer of $^{252}$Cf.

The raise of the experimental spectrum in the region of 45-th channel is caused by the pulses overlapping from the α-activity of the layers. The zero of the amplitude scale is on the 26-th channel. The raise in the range of 26-30 channels corresponds to the part of fragments not escaping into the gas.

The ratio of the energies of light and heavy fragments is $\approx (1.35 \div 1.40)$. On Fig.2 the maximum of light fragments is on $\approx 230$ and of heavy ones on 170 channels. Their average path lengths are $\approx 27$ mm and 21 mm at the pressure of 1 atm and are 5 times less at 5 atm, i.e. $\approx 5$ mm and 4 mm correspondingly. If the escape angle is $\theta < \arccos(d/R) \approx 78^0$, the light fragment loses the part of energy in the bottom of the chamber and, as a consequence, the pulse amplitude from it will move to the 120-200 channels, while the heavy fragment – from $\approx 75^0$ and the pulse amplitude from it will move to 120-150 channels.

The absolute velocity of fission in chamber was defined in accordance with technique where there were introduced the notions of the operational registration of effectiveness of fission fragments $\mathcal{E}_{oper}$ and the true registration of efficiency $\mathcal{E}_{true}$. It has been shown on practice that it is convenient to determine the effectiveness for different channels by referring to the position of the half of decreasing, which is defined as the maximum of the positive derivative of the basic distribution of the fragments' spectrum (*max +*). In this case (*max +*) corresponds to the 114-th channel. Using calculated spectrum, one can find the number of readings of pulses in the fragments' spectrum in the middle of the range *b* between (*max +*) and the zero (P=26 channel) of the amplitude scale:

$$b = 0.5 \times ((max\ +) - P), \quad (3)$$

i.e. in this case at the 70-th channel $C_{70}$, the true fragments' reading on the left and right from the 70-th channel $C_{left}$ and $C_{right}$ and to determine the true efficiency of the registration:

$$\mathcal{E}_{true} = 1 - (C_{left} / C_{right}). \quad (4)$$

The operational efficiency $\mathcal{E}_{oper}$ is calculated by the rates of fragments readings in the 70-th channel $C_{70}$ with suggestion that the rates will be the same also in channels on the left (i.e. $\mathcal{E}_{oper}$ is a rectangle area with the height $C_{70}$ and the length 70-26=44 channels), divided by the square $C_{right}$:

$$\mathcal{E}_{oper} = 1 - (C_{70} - (70 - 26)) / C_{right}. \quad (5)$$

Calculations are made with different thicknesses of layers of fission element and for different values of the last deformation (the 'sphericity' of the bottom) and for the overshot of the gaseous layer in the range of values (0.2÷1.5) mm.

It is turned out that the ratio $y = (1 - \mathcal{E}_{true}) / (1 - \mathcal{E}_{oper})$ is rather smooth function of the value of the "width" of the fragment distribution $\Delta = (max\ -) / (max\ +)$, where (*max -*) is the position of the middle of decreasing, i.e. the maximum of the negative derivative of the basic distribution of the fragments' spectrum. On the Fig.2 (*max -*) is on ≈ 250 channel.

By determining from the experimental spectrum the operational efficiency $\mathcal{E}_{oper}$ and the "width" $\Delta$ of the distribution of the fragments, one finds the value of *y* and $\mathcal{E}_{true}$:

$$\mathcal{E}_{true} = 1 - y \times (1 - \mathcal{E}_{oper}), \quad (6)$$

where $\mathcal{E}_{oper}$ can be determined from the experimental spectrum (Fig. 2).

To revise the full and partial neutron cross-sections with errors ≤5% and their integral characteristics (the resonance parameters, the factors of the resonance interlock and the Doppler's coefficients with errors ≤15%) of the constructive reactor materials Ti, Mn, Cr, Fe, Ni, Ta, Mo, the necessary experiments by the time-of-flight technique at the neutron source RADEX and LSDS-100 with TOF of INR RAS have been made. It is expected to make measurements of the time-of-flight spectrums for the samples of filters and radiators of thickness 40 mm, 20 mm, 5 mm, 2 mm, 1 mm on the path length of 50 m of the neutron pulsed source RADEX on the experimental installation (the resonance cross-section of transmission) with the fast 8-sectional liquid-based (n,γ)-detector and the high efficient neutron He-3 detector. The experimental time-of-flight spectrums will be stored in four independent measuring modules with different durations of the time channels τ=0,20 μs, 1 μs, 2 μs, 4 μs in the group energetic intervals of the constant system BNAB in the energy range from the thermal neutrons 0.1 eV up to the fast ones 200 keV.

To achieve the required accuracy of the being obtained nuclear-physics values of the mentioned above materials, it is necessary to have the stable operation of the proton accelerator at the experimental complex during 14 operating shifts with the proton energy 209 MeV, the pulse current 16 mA, the time duration of the neutron flashes (5-20) μs and the frequency of the neutron flashes 50 Hz.

In case of the considerable losses during the beam transport, it is possible to use the usual 7 mA beam current by turning off of the first and the second bunches.

There have been determined from the experimental spectrums the cross-sections, transmissions, self-indications and other integral characteristics, for example, the factors of the resonance interlock with the error less than 5%. Besides of that, the deep interference dips let us estimate the energy dependence of the cross-section of the ($n,e$) scattering.